\documentclass{article}

\usepackage{amsmath,epsfig}
\usepackage{amssymb,amsfonts}
\usepackage{latexsym}

\relax

\usepackage{graphicx}

\newcommand{\AP}{\alpha^{\prime}}
\newcommand{\pd}{\partial}

\newcommand{\Fc}{\mathcal{F}}
\newcommand{\Gc}{\mathcal{G}}

\begin{document}

\title{Multi-scalar field cosmology from SFT:\\an exactly solvable
approximation}

\author{F.~Galli\footnote{Aspirant FWO-Vlaanderen, e-mail:
\texttt{fgalli@tena4.vub.ac.be}} ~and A.~S.~Koshelev\footnote{Postdoctoral
researcher of
FWO-Vlaanderen, e-mail: \texttt{alexey.koshelev@vub.ac.be} 
}\\
~~\\
\textit{\small Theoretische
Natuurkunde, Vrije Universiteit Brussel
and}\\ \textit{\small The International Solvay Institutes,}\\ \textit{\small Pleinlaan 2, B-1050
Brussels,
Belgium}}

\date{~}

\maketitle
\vfill
\begin{abstract}
We consider the appearance of multiple scalar fields in SFT inspired non-local
models with a single scalar field at late times. In this regime all the scalar
fields are free. This system minimally coupled to gravity can be analyzed
approximately or numerically. The main result of this note is the introduction
of an exactly solvable model which obeys an exact solution in the
cosmological context for the Friedmann equations and that reproduces
the behavior expected from SFT in the asymptotic regime.
 Different applications of such a potential to
multi-field cosmological models are discussed.
\end{abstract}
\vfill



\section{Introduction}

In this note we briefly review a new class of
cosmological models based on  string field
theory (SFT) (for details see reviews~\cite{review-sft}) and 
$p$-adic string theory \cite{padic}. We introduce an exactly solvable
model reproducing the  behavior of the initial model in  the asymptotic regime (see \cite{IA1,Koshelev07,AKd,KVlast} and references therein for
a more detailed analysis on the subject).
It is known that  SFT and  $p$-adic string theory are
UV-complete.  One can therefore expect that the resulting (effective) models
should be free of pathologies. Furthermore, models originating from 
SFT exhibit one general non-standard property: they have terms
with infinitely many derivatives, i.e. non-local terms. Higher
derivative terms usually produce the well known Ostrogradski
instability \cite{ostrogradski}.
However, the Ostrogradski result is related to derivatives order higher than two but
finite. In the case of infinitely many
derivatives it is possible that instabilities do not appear.

Contemporary cosmological
observational data~\cite{data,Komatsu} strongly support that the
present Universe exhibits an accelerated expansion providing thereby an
evidence for a dominating DE component~\cite{review-de}. Recent results of
WMAP~\cite{Komatsu} together
with the data on Ia supernovae give the following bounds for the DE
state parameter $ w_{\text{DE}}=-1.02^{+0.14}_{-0.16} $. Note that the
present cosmological observations do not exclude an evolving DE state
parameter $w$.
Non-local models of the type obtained from SFT may have effective phantom
behavior and can therefore be 
interesting for the present cosmology. To construct a stable model with
$w<-1$ one should construct from the fundamental theory, which is stable and
admits quantization, an effective theory with the Null Energy Condition
(NEC) violation.
This is a hint towards
SFT inspired cosmological models.


\section{Model setup}

We are interested in non-local models arising from SFT in large-time regime . Our starting point is the action
\begin{equation}
S=\int d^4x\sqrt{-g}\left(\frac{R}{16\pi
G_N}+\frac1{g_o^2}\left(-\frac12\pd_\mu T\pd^\mu
T+\frac1{2\AP}T^2-\frac1{\AP}{v(\bar T)}\right)-\Lambda'\right) \, .
\label{action_model_pre}
\end{equation}
We work in $1+3$ dimensions with
 signature $(-,+,+,+)$, the coordinates are denoted by
Greek indices $\mu,\nu,\dots$, running from 0 to 3.
$G_N$ is the Newtonian constant, $8\pi G_N=1/M_P^2$, where
$M_P$ is the Planck mass, $\AP$ is the string length squared (we do not
assume $\AP=1/M_P^2$), $g_o$ is the open string coupling constant and  $\Lambda'$ is a constant term.
 $\bar T=\Gc(\AP\Box)T$, with
$
\Box=D^{\mu}\pd_{\mu}=\frac1{\sqrt{-g}}\partial_{\mu}\sqrt{-g}g^{\mu\nu}
\partial_{\nu}
$ and $D_\mu $ is the covariant
derivative,  $T$ is a scalar field primarily associated with the open string tachyon.
The function $\Gc(\AP\Box)$ can be non-polynomial, thus producing a manifest
non-locality.
Fields are dimensionless while $[g_o]=\text{length}$. $v(\bar T)$ is the open
string tachyon self-interaction. It does not have a term quadratic in $T$.
Factor $1/\AP$ in front of the tachyon
potential looks unusual and can be easily removed by a rescaling of
fields. For our purposes it is convenient to keep all the fields dimensionless.

Such a  four-dimensional action is motivated
by string field theory ( see \cite{Koshelev07} for details ).
In SFT one has $\Gc(\AP\Box)=e^{-\frac{\beta}2\AP\Box}$,
 where $\beta$ is a parameter determined exclusively by the conformal field theory of
the string, but we keep this function general. We stress once again that appearance of non-localities is a general
feature of SFT based models and it is exactly the feature that we are going to
explore here\footnote{The appearance of higher derivatives is not an exclusive
feature of this theory. Non-commutative theories, for
instance, also have higher derivative, but these non-local structures are very
different.}.

Introducing the field $T_b=\bar T$ and dimensionless coordinates
$\bar{x}_\mu=x_\mu/\sqrt{\AP}$, we can rewrite the
above action   as
\begin{equation}
S=\int d^4x\sqrt{-g}\left(\frac{R}{16\pi
G_N}+\frac1{g_o^2}\left(-\frac12\pd_\mu \tilde T_b\pd^\mu
\tilde T_b+\frac1{2}\tilde T_b^2-{v(T_b)}\right)-\Lambda'\right). \label{action_model}
\end{equation}
where we omit bars for simplicity and set hereafter $\AP=1$.
We emphasize that the potential of the field $T_b$ is
$V=-\frac1{2}T_b^2+v(T_b)$.  Assuming an extremum of the potential $V$  exists, one can
linearize the theory around it using $T_b=T_0-\tau$. As a result one obtains
$$V=-\frac12\tau^2+\frac{v(T_0)''}2\tau^2+V(T_0).$$
The action (\ref{action_model}), linearized around the extremum of the potential, can be written as
\begin{equation}
S=\int d^4x\sqrt{-g}\left(\frac{R}{16\pi
G_N}+\frac1{2g_o^2}\tau\Fc(\Box)\tau-\Lambda\right),
\label{action_model2}
\end{equation}
where $ \Fc(\Box)=(\Box+1)\Gc^{-2}(\Box)-m^2 $ with  $m^2\equiv v(T_0)''$ and $\Lambda=\Lambda'+\frac{V(T_0)}{g_o^2}$.
 In SFT one would have 
$ \Fc_{\text{SFT}}(\Box)=(\Box+1)e^{\beta\Box}-m^2$. $\Fc$ is in fact the inverse propagator and it is natural to expect $\beta<0$
in the SFT case corresponding to  convergent propagator at large momenta. From
this SFT example we can draw a very important lesson. Assume $m=0$. Then
$\Fc_{\text{SFT}}=(\Box+1)e^{\beta\Box}$ and the non-locality does not show up
at all. Indeed, the poles of the propagator do not feel the exponent. Another
way of thinking is that the exponential factor can be eliminated by a field
redefinition. The situation is dramatically different for $m\neq0$. The
propagator $1/\Fc_{\text{SFT}}$ has infinitely many poles. This is a
manifestation of the non-locality. Also note that the physics would be totally
different for full function $\Fc$ and its truncated series expansion because the
pole structure may get modified considerably.

A very important role is played by the spectrum of the theory, determined by the equation
\begin{equation}
    \Fc(J)=(J+1)\Gc^{-2}(J)-m^2=0.\label{cheq}
\end{equation}
We call it \textit{characteristic} equation. It can be an algebraic or
a transcendental. We consider  $\Fc(\Box)$ of  general form, with the only assumption that all roots are simple. The
analyticity of the function $\Fc$ on the complex plane is also important for representing $\Fc$ by the convergent series
expansion:
\begin{equation}
\Fc=\sum\limits_{n=0}^{\infty}f_n\Box^n \, ,\quad f_n\in\mathbb{R}.
\end{equation}
Reality of coefficients is required by the hermiticity of the Lagrangian.
Strictly speaking even the analyticity requirement can be weakened, but in this
case one has to be careful with poles and the corresponding convergence domain of the series.

Classical solutions of the equation of motion were
studied and analyzed in \cite{Koshelev07}. The key point in the  analysis is the  fact that action (\ref{action_model2})
is fully equivalent to the action 
\begin{equation}
S_{local}=\int d^4x\sqrt{-g}\left(\frac{ R}{16\pi
G_N}-\frac{1}{g_o^2}\sum_i\frac{\Fc'({J}_i)}{2}\left(g^{\mu\nu}
\pd_\mu\tau_i\pd_\nu\tau_i
+{J}_i\tau_i^2\right)-\Lambda\right) \label{action_model_local}
\end{equation}
with many free local scalar fields. Here $J_i$ are roots of the characteristic equation (\ref{cheq}) and there are
as many scalar fields as  roots of the characteristic equation.
The details of the equivalence statement can be found in
\cite{Koshelev07,KVlast,nonloc-loc}.

It is important  that the roots $J_i$ as
well as the coefficients $\Fc'(J_i)$ can be complex. This do not represent  a problem,  
since all the local scalar fields are just mathematical functions without physical interpretation.
What is important is that the original field $\tau$ must
be real since it represents a physical excitation. This is simple to
achieve. The roots $J_i$ are either real or complex conjugate. Complex
conjugate $J_i$ would yield complex conjugate (up to an overall constant factor)
solutions for $\tau_i$. Thus, making the linear combination of all $\tau_i$ real  is just a
matter of choosing the integration constants appropriately. Finally, note that  the 
coefficients $\Fc'(J_i)$ can be eliminated  by a field rescaling.

In \cite{KVlast} it was proven  that cosmological perturbations in the free
theory with one non-local scalar field
(\ref{action_model2}) and in the corresponding local theory
(\ref{action_model_local}) with many scalar field are equivalent as well.
Moreover, several examples of
evolution and perturbations were developed. The difficulty here, however, is
that not many exactly solvable models with multiple scalar fields are known. Studying perturbations without having
an exact solution is not an easy task. Of course, there is no need to have infinitely many scalar
fields. On the contrary, on can set almost all of them to zero by
choosing trivial integration constants. The point is that the models for 
which exact solutions are known are mostly models with one single field. Reducing 
the action  (\ref{action_model_local}) to a single local field completely
misses the rich and non-trivial structure coming from SFT. The most
intriguing case of complex masses requires at least two fields. In fact, we need
the original function $\tau$ to be real and thus one cannot keep only one
complex field without its complex conjugate.


\section{Exactly solvable model}

An exact analytic solution for equations of motion following from the action
(\ref{action_model_local}) (and furthermore (\ref{action_model2})) is not
known. However, it is much more transparent to work with exact solutions rather
than with asymptotics when one wants to study cosmological perturbations. There
is a chance to modify the potential such that: first, the model becomes exactly
solvable and, second,  all the new terms vanish
rapidly in the regime of interest, so that the previous picture is restored. Furthermore an exactly
solvable model in cosmology has its own value just because not so many exactly
solvable models are known. In our particular case we deal with many scalar
fields and this complicates the problem. Moreover we have complex coefficients
in the Lagrangian and this is an unexplored problem.

We consider the following modified action
\begin{equation}
\begin{split}
 S=&\int d^4x\sqrt{-g}\Biggl[\frac{ R}{16\pi
G_N}+\frac{1}{g_o^2}\Biggl(-\sum_i\frac{\Fc'({J}_i)}{2} \Bigl(
g^{\mu\nu}(\pd_\mu\tau_{i+}\pd_\nu\tau_{i+} +\pd_\mu\tau_{i-}\pd_\nu\tau_{i-})+\\ 
+&{J}_i(\tau_{i+}^2+\tau_{i-}^2)\Bigr)
-\frac{3\pi G}2\left(\sum_{i}\Fc'({J}_i)( \alpha_{i+}\tau_{i+}^2
  +\alpha_{i-}\tau_{i-}^2)\right)^2 + c.c.\Biggr)-\Lambda\Biggr]\, ,
\end{split} \label{action_model_localexactpm}
\end{equation}
where $H_0=\sqrt{\frac{8\pi  G_N\Lambda}3}$,  $G \equiv G_{N} / g_{o}^{2}$ is
a   dimensionless analog of the Newton constant and $\alpha_{i\pm}$ are  the two solutions\footnote{In the special
    case  $\frac{4J}{9H_{0}^{2}}=1$ the two solutions coincide and we are left with just one
    scalar field.}of  $J_i=-\alpha_i(\alpha_i+3H_0)$. 
To simplify the notation, we introduce new indeces $P,Q,R\dots$
taking the degeneracy in $\alpha_{i}$ into account. Namely, we consider $P,Q,R\dots =
i_{+} , i_{-} ,j_{+}\dots$, bearing in mind that one can now have   $J_{P} =J_{Q}$ for some values of $P$ and $Q$. With this piece
of notation the modified action \eqref{action_model_localexactpm} takes the form
\begin{equation}
 S\!=\! \!\int d^4x\sqrt{-g}\Bigg[\frac{ R}{16\pi
G_N}+\frac{1}{g_o^2}\Biggl(-\tilde{\sum_{P} } \frac{\Fc'({J}_P)}{2}\left(g^{\mu\nu}\pd_\mu\tau_P\pd_\nu\tau_P
+{J}_P\tau_P^2\right)
-\frac{3\pi G}2\left(\tilde{\sum_{P}} \Fc'({J}_P)\alpha_P\tau_P^2\right)^2\Biggr)-\Lambda\Biggr]
\label{action_model_localexact}
\end{equation}
where $\tilde\sum_{P}$ indicates that complex conjugate quantities are
  included in the sum\footnote{For example $\tilde\sum_{P}   J_{P} \tau^{2}_{P} =
    \sum_{P}  \left(  J_{P} \tau^{2}_{P} + J^{*}_{P} \tau^{2*}_{P}
    \right)$. }.
    
We now specialize to the case of  spatially flat Friedmann--Robertson--Walker (FRW) Universe, with metric 
\begin{equation}
\label{mFr} ds^2={}-dt^2+a^2(t)\left(dx_1^2+dx_2^2+dx_3^2\right) \, , 
\end{equation}
where $a(t)$ is the scale factor and $t$ is the cosmic time. The Hubble
parameter, as usual,  is  $H=\dot a/a$ and the dot here and hereafter in this paper denotes a
derivative with respect to  $t$. Background solutions for $\tau$ are therefore assumed  to be spacially homogeneous
as well. The local action \eqref{action_model_localexact} now admits an exact solution to its equations of motion,
which are
\begin{equation}
\label{FrEOMgFRWex}
\begin{split}
3H^2&=4\pi G\left(\tilde{\sum_P}\Fc'({J}_P)\left(\dot\tau_P^2
+{J}_P\tau_P^2\right)+3\pi G\left(\tilde{\sum_{P}}\Fc'({J}_P)\alpha_P\tau_P^2\right)^2\right)+8\pi G_N\Lambda\, ,\\
\dot H&=-4\pi G
\tilde{\sum_P}\Fc'({J}_P)\dot\tau_P^2\, ,\\
\end{split}
\end{equation}
and
\begin{equation}
\ddot\tau_P+3H\dot\tau_P+J_P\tau_P+{6\pi
G}\alpha_P\tau_P\tilde{\sum_{Q}}\Fc'({J}_Q)\alpha_Q\tau_Q^2=0,~~\text{~for all}~P\, .
\label{FrEOMtauFRWex}
\end{equation}
One can explicitly check that there is the following solution\footnote{Similar
modification of the action was considered in \cite{ajv4} and the analysis in the
case of real roots was performed.}
\begin{equation}
\begin{split}
\tau_P&=\tau_{P0}e^{\alpha_P t}\, ,\\
H&=H_0-2\pi G\tilde{\sum_P}\Fc'({J}_P)\alpha_P\tau_{P0}^2e^{2\alpha_P t}\, .
\end{split}\label{action_model_localexactsol}
\end{equation}

This solution is valid for any number of fields (including single field model)
and for any values of parameters $\alpha_P$ (i.e. real, complex,
etc.).
Moreover, we see that if $\text{Re}(\alpha_P)<0$ for all $P$ then the quartic
term in the scalar fields potential vanishes and we are left with free fields.
Thus for large times the model (\ref{action_model_local}) is restored and we can
speak about the asymptotic regime of SFT based models. Stability of the
 solution deserves a deeper analysis and methods used in \cite{abv} can
be applied.


\section{Cosmological perturbations}
 
In this section we want to present the equations describing cosmological perturbations
for the exact solution  \eqref{action_model_localexactsol}.

For the action \eqref{action_model_localexact}  one has the following
perturbation equations for many scalar fields: 
\begin{align}
&\ddot\zeta_{PQ} + \dot\zeta_{PQ}\left( 3 H +
  \frac{\ddot\tau_{P}}{\dot\tau_{P}} +
  \frac{\ddot\tau_{Q}}{\dot\tau_{Q}} \right)  + \zeta_{PQ} \left(
  -3 \dot H + \frac{k^2}{a^2} \right) =  \left[ \frac{\tau_{P}}{\dot\tau_{P}} \left( J_{P}  + 6 \pi G \alpha_{P}
    \tilde{\sum_{R} }\Fc'({J}_R)\alpha_{R} \tau^{2}_{R} \right)  \right. \nonumber \\
& \left. -  \frac{\tau_{Q}}{\dot\tau_{Q}} \left( J_{Q}  + 6 \pi G \alpha_{Q}
    \tilde{\sum_{R}} \Fc'({J}_R) \alpha_{R} \tau^{2}_{R} \right)   \right]\left( \tilde{\sum_{S}}
  \Fc'({J}_S)\frac{\dot \tau^{2}_{S}}{ \rho + p }\left( \dot\zeta_{PS} +
   \dot\zeta_{QS}\right) + \frac{2 \varepsilon}{1+w}\right) +
\nonumber \\
& + 12 \pi G \tilde{\sum_{R} }\Fc'({J}_R)\alpha_{R} \tau_{R} \dot\tau_{R} \left( \alpha_{P}
  \frac{\tau_{P}}{\dot\tau_{P}} \zeta_{PR}  -  \alpha_{Q}
  \frac{\tau_{Q}}{\dot\tau_{Q}} \zeta_{QR}\right) \label{perturb_zeta}
\, ,
\end{align}
\begin{align}
&\ddot\varepsilon + \dot\varepsilon H \left( 2 - 6 w + 3 c^{2}_{s}
\right) + \varepsilon \left( \dot H (1 - 3 w) - 15 w H^2 + 9 H^2
  c^{2}_{s} + \frac{k^2}{a^2} \right) = \nonumber \\
=& \frac{k^2}{a^2} \frac{1}{\rho} \frac{2}{\rho + p } \tilde{\sum_{R,S}}
\left(J _{R }  + 6 \pi G\alpha_{R}\tilde{\sum_{P}} \Fc'({J}_P)\alpha_{P} \tau^{2}_{P}\right)
\Fc'({J}_R) \Fc'({J}_S)\tau_{R}\dot\tau_{R} \dot\tau^{2}_{S} \zeta_{RS} \label{perturb_eps}
\, .
\end{align}

Here $\zeta_{PQ}=\frac{\delta\tau_P}{\dot\tau_P}-\frac{\delta\tau_Q}{\dot\tau_Q}$ is
the gauge invariant variable for the scalar fields perturbation and
$\varepsilon$ is the
gauge invariant total energy density perturbation and $k$ is the
  comoving wavenumber (see, for instance,
\cite{bardeen,Mukhanov,hwangnoh,KVlast} for a derivation of perturbation
equations and details). The collective energy density and pressure
are  $$\rho =
\frac{1}{2}\tilde\sum_{P}\Fc'({J}_P)\dot\tau^{2}_{P} +  V\, ,  \qquad p =
\frac{1}{2}\tilde\sum_{P}\Fc'({J}_P)\dot\tau^{2}_{P} -  V\, ,$$ with  $$V=
\frac{1}{2}\tilde\sum_{P}\Fc'({J}_P) J_{P}\tau^{2}_{P}  + \frac{3 \pi
G}{2}\left(\tilde\sum_{P}\Fc'({J}_P) \alpha_{P}  \tau^{2}_{P} \right)^2\, .$$   We
also
introduced the following notation   $w \equiv p/\rho$ for  the equation of state
  parameter and  $c_{s}^{2} \equiv \dot p / \dot\rho$ for the speed of
sound .
Using the explicit solution  \eqref{action_model_localexactsol}  one
finds
\begin{equation*}
\begin{split}
\rho &=\frac{1}{2} \tilde{\sum_{P}}\Fc'({J}_P)\tau_{0P}^{2}e^{2 \alpha_{P} t}\left(
  \alpha_{P}^{2} + J_{P} \right) +\frac{3 \pi G}{2}\left(
 \tilde{\sum_{P}}\Fc'({J}_P) \alpha_{P}\tau_{0P}^{2}e^{2 \alpha_{Pi} t} \right)^2
+g^{2}_{0}\Lambda = \frac{3}{8\pi G} H^2 \, ,\\
p &=\frac{1}{2} \tilde{\sum_{P}}\Fc'({J}_P)\tau_{0P}^{2}e^{2 \alpha_{P} t}\left(
  \alpha_{P}^{2} - J_{P} \right) -\frac{3 \pi G}{2}\left( \tilde{\sum_{P}}\Fc'({J}_P)\alpha_{P}\tau_{0P}^{2}e^{2 \alpha_{P} t} \right)^2  - g^{2}_{0}\Lambda
= -  \frac{\dot H}{4 \pi G } - \frac{3}{8\pi G} H^2\, ,
\\
w &=\frac{ \tilde\sum_{P}\Fc'({J}_P)\tau_{0P}^{2}e^{2 \alpha_{P} t}\left(
  \alpha_{P}^{2} - J_{P}  -3 \pi G \alpha_{P}
  \tilde\sum_{Q}\Fc'({J}_Q)\alpha_{Q}\tau_{0Q}^{2}e^{2 \alpha_{Q} t} \right)  -
g^{2}_{0}\Lambda }{ \tilde\sum_{P}\Fc'({J}_P)\tau_{0P}^{2}e^{2 \alpha_{P} t}\left(
  \alpha_{P}^{2} + J_{P}  + 3 \pi G \alpha_{P}
  \tilde\sum_{Q}\Fc'({J}_Q)\alpha_{Q}\tau_{0Q}^{2}e^{2 \alpha_{Q} t} \right)  +
g^{2}_{0}\Lambda } = -1 - \frac{2}{3}\frac{\dot H}{H^2} \, ,
\\
c_{s}^{2}  &= \frac{ \tilde\sum_{P} \Fc'({J}_P)\alpha_{P} \tau^{2}_{0P}  e^{2 \alpha_{P}
    t}   \left( \alpha_{P}^{2} - J_{P} - 6 \pi G \tilde\sum_{Q} \Fc'({J}_Q)\alpha^{2}_{Q} \tau^{2}_{0Q} e^{2 \alpha_{Q} t}
 \right) }{\tilde\sum_{P} \Fc'({J}_P)\alpha_{P} \tau^{2}_{0P} e^{2 \alpha_{P} t}
 \left( \alpha_{P}^{2} +  J_{P} + 6 \pi G \tilde\sum_{Q} \Fc'({J}_Q)\alpha^{2}_{Q}
   \tau^{2}_{0Q} e^{2 \alpha_{Q} t} \right)} = -1 - \frac{\ddot H }{3
 \dot H H } \, .
\end{split}
\end{equation*}
Substituting the explicit solution for $\tau_P$  in  \eqref{perturb_zeta} one
notices that the term given by  the second derivative of the potential in the
r.h.s. cancels with the one proportional to
$\dot H$ in the l.h.s.
Equation  \eqref{perturb_eps}
evaluated on the
exact solution gets simplified because the contribution 
coming from the quartic
term  in the potential drops in this case.
One can express these two equations in terms of $H$ and its
derivatives and also eliminate $J_P$  form all these expression by
replacing it with the combination  $-\alpha_P ( \alpha_P + 3H_0
)$.  All the $ H_{0}$ contributions coming form this substitution simply
drops because of the symmetry properties of the expressions involving $J_P$.
\begin{align}
\ddot\zeta_{PQ} & + \dot\zeta_{PQ}\left( 3 H +
  \alpha_{P} +
  \alpha_{Q} \right)  + \zeta_{PQ} \frac{k^2}{a^2}   = \nonumber \\
&= \frac{1}{\dot H} \left(
  \alpha_{P}  -  \alpha_{Q}  \right)\left( 4 \pi G \tilde{\sum_{R}} \Fc'({J}_R)
 \tau^{2}_{0R} \alpha^{2}_{R} e^{2 \alpha_{R}t}\left( \dot\zeta_{PR} +
    \dot\zeta_{QR}\right) + 3 H^2 \varepsilon\right)\, ,
\end{align}
\begin{align}
\ddot\varepsilon +& \dot\varepsilon \left( 5 H  + 4 \frac{\dot H}{H}
  - \frac{\ddot H}{\dot H} \right) + \varepsilon \left( 6 H^2  + 14
  \dot H + 2 \frac{\dot H^2 }{H^2}   - 3 H  \frac{ \ddot H}{\dot H}+
  \frac{k^2}{a^2} \right) =  \nonumber \\
& =\frac{k^2}{a^2} \frac{4}{3} \frac{(4 \pi G)^2}{\dot H H^2 } \tilde{\sum_{R,S}}
  \Fc'({J}_R) \Fc'({J}_S)\alpha_{R} \alpha^{2}_{R} \alpha^{2}_{S} \tau^{2}_{0R}\tau^{2}_{0S}
 e^{2\alpha_{R}t} e^{2\alpha_{S}t} \zeta_{RS}\, . 
\end{align}

The two above equations form the main result of the present note.
Analysis of these equations is a very important problem currently being studied.

The example of perturbations with complex roots in the original
linearized action (\ref{action_model_local}) was carried out in \cite{Klast}.
Linear perturbations in such a configurations can be confined thus
not destroying the system itself. This result is not obvious
and it supports the claim that the SFT based models are stable.
The case of complex $J_P$ has never
been studied in general and deserves deeper investigation.  Analysis of
perturbations with complex roots in the exactly solvable model
will be addressed in  a forthcoming publications.


\section*{Acknowledgments}
A.K. would like to thank the organisers of the ``Bogolyubov 2009''
conference for the opportunity to present this work and for
creating a very stimulating environment for scientific discussions.
The authors are grateful to I.Ya.~Aref'eva, 
F.~Bezrukov, B.~Craps, B.~Dragovich, G.~Dvali,
V.~Mu\-kha\-nov, and  S.Yu.~Vernov for useful comments and stimulating
discussions. This work is
supported in part by the Belgian
Federal Science Policy Office
through the Interuniversity Attraction Poles IAP VI/11, the European
Commission FP6 RTN programme MRTN-CT-2004-005104 and by FWO-Vlaanderen
through the project G.0428.06.
A.K. is
supported in part by RFBR grant 08-01-00798 and state contract of
Russian Federal Agency for Science and Innovations 02.740.11.5057.

\end{document}